\documentclass[%
 reprint,
 amsmath,amssymb,
 aps,
]{revtex4-2}

\usepackage{graphicx}
\usepackage{dcolumn}
\usepackage{bm}

\begin{document}

\preprint{APS/123-QED}

\title{Interaction uncertainty in financial networks}

\author{Andrea Auconi}

\affiliation{%
Ca’ Foscari University of Venice, DSMN , via Torino 155, 30172 Mestre (Venice), Italy
}%


\date{\today}

\begin{abstract}
A minimal stochastic dynamical model of the interbank network is introduced, with linear interactions mediated by an integral of recent variations. Defining stress as the variance over the banks' states, the interaction correction to the stress expectation is derived and studied on the short-medium timescale in an expansion. It is shown that, while different interaction matrices can amplify or absorb fluctuations, on average interactions increase the stress expectation. More in general, this analytical framework enables to estimate the impact of uncertainty about financial exposures, and to draw conclusions about the importance of disclosure.
\end{abstract}



\maketitle


\paragraph*{Introduction.} 

The stability of the financial system is at the basis of a functioning economy, where banks enable business by lending and investing under a solid regulatory framework \cite{hull2003options}.
Investment and risk hedging are such that banks form a network of assets and liabilities whose market values are affected by interest rates, credit risk, volatility, and more \cite{piterbarg2010funding,brigo2013counterparty}.
This interconnectedness creates the conditions for contagion and systemic risk, where shocks on individual banks get amplified by the network dynamics \cite{gai2011complexity,glasserman2016contagion,caccioli2018network,bardoscia2021physics}.

The study of the Jacobian around fixed points of dynamical systems, where eigenvalues determine the stability to perturbations, has widely been adopted to characterize the long-term behavior of complex networks \cite{may1972will,haldane2011systemic,barzel2013universality,bardoscia2017pathways,meena2023emergent,bontorin2023multi}.
For the shorter timescales, on the other hand, the particular realization of the fluctuations matters \cite{risken1996fokker,shreve2004stochastic} and characterizations are typically given in terms of probability functionals, e.g.  derivatives prices \cite{shreve2004stochastic,brigo2006interest}, response coefficients \cite{marconi2008fluctuation,dechant2020fluctuation,auconi2021fluctuation}, entropy production \cite{parrondo2015thermodynamics,horowitz2020thermodynamic,ito2023geometric,auconi2019information}.
Motivated by the perspective of financial regulators, a stress expectation is here considered as probability functional.

In this Letter, we introduce a minimal stochastic dynamical model of how exposures propagate fluctuations through the financial network. Abstracting from the complexity of trade types, the model employs an interaction matrix acting on an integral of recent variations.
We analytically derive the interaction matrix properties which determine if, on the short-medium term, the stress is absorbed or rather amplified. We also derive the impact on the stress expectation of the uncertainty about such interaction matrix, and it will motivate the importance of exposures disclosure.

\paragraph*{The interbank model.}
Let us consider a network of $N$ banks and a corresponding set of variables $x(t) \equiv \{x_i (t)\}_{i=1,...,N}$ representing the state at time $t$ of the individual banks. These variables can model (the logarithm of) the book value, stock price, asset swap spreads, or some synthetic measure of financial health.
In a nutshell, the book value of a bank $i$ fluctuates over time due to price variations of its external assets, and these fluctuations will then affect the book value of another bank $j$ if it is a counterparty of $i$ for example by holding its bonds. This propagation of fluctuations is however not immediate because bonds are market traded and cannot follow the book value variations immediately, also because earnings and other impactful news are only disclosed periodically.
The minimal interbank network stochastic dynamics model reads
\begin{equation}\label{model}
    dx_i = \sigma dW_i + \gamma M_{ij} h^j dt,
\end{equation}
where the adapted stochastic process $dh_i = -\beta h_i dt + \sqrt{\beta}\, dx_i$ is the recent variation of $x_i$ over the timescale $\beta^{-1}$, which models the finite speed of the market reaction and how it mediates the linear interactions between banks.

The uncertainty of banks' profits due to market fluctuations and non-modelled factors is described using uncorrelated standard Brownian motion increments, $\{dW_i (t)\}_{i=1,...,N}$ in the It\^o notation \cite{shreve2004stochastic}. These follow the standard covariance $\mathbb{E}\,dW_i(t)dW_j(t')=\delta_{ij}\delta_{tt'}dt$, where $\int_{0}^{T} \delta_{tt'} f(t') \equiv f(t) \mathbb{I}_{0<t<T}$ defines the Dirac delta for a smooth function $f(t)$, $\mathbb{I}$ is the indicator function, and $\delta_{ij}$ the Kronecker delta. Sums are implicit following Einstein convention, $M_{ij} h^j\equiv \sum_j M_{ij} h_j$, and we consider the processes to start at $x_i(0)=0$ and $h_i(0)=0$ for all $i$s.
$\hat{M}\equiv \{M_{ij}\}_{i,j=1,...,N}$ is the interaction matrix, and $\gamma$ the interaction strength.
This model is meant as a coarse-grained description of how financial exposures drive the propagation of individual nodes' fluctuations through the network, without modelling the complexity of trade types. Note that not only standard exposures in terms of trades' profit-and-loss, but also changes in calibration models can propagate fluctuations \cite{barucca2020network}.
To focus on the effect of interactions we now keep the volatility $\sigma > 0$ to be constant over time, and identical for all nodes. Later the model is extended with stochastic volatility.



The recent variation $h_i\equiv h_i(t)$ follows an Ornstein-Uhlenbeck process whose formal solution in terms of the matrix exponential is, see \cite{risken1996fokker},
\begin{equation}\label{history}
    h_i(t) = \sigma \sqrt{\beta} \int_0^t \left( e^{-\hat{A}(t-t')} \right)_{ij} dW^j (t'),
\end{equation}
where we defined $\hat{A}\equiv \beta \hat{\delta} -\sqrt{\beta} \gamma \hat{M}$, and $\hat{\delta}$ denotes the identity matrix. We note that in the absence of interactions, namely $\gamma=0$, the recent variation is stationary with zero mean and variance independent of $\beta$, $\lim_{t\rightarrow\infty} \mathbb{E}\left[h_i^2(t)\right] \big{|}_{\gamma=0} = \sigma^2/2$.
Integrating Eq. \eqref{model} and using Eq. \eqref{history} we obtain
\begin{equation}\label{formal solution}
    x_i (t) = \sigma W_i(t) +\sigma \sqrt{\beta} \gamma S_i(t),    
\end{equation}
where $W_i(t)=\int_0^t dW_i(t')$ is standard Brownian motion and $S_i(t)$ is the stochastic integral
\begin{equation}
    S_i(t) = \int_0^t dt' \int_0^{t'} M_{ij} \left( e^{-\hat{A}(t'-t'')}\right)^{jk} dW_k (t''),
\end{equation}
which sums the propagation onto node $i$ of the fluctuations in all nodes $k$ during the time interval $[0,t)$.


\paragraph*{The stress observable.} 

Let us consider the sample variance of the banks states as a quantifier of stress in financial networks. For a particular realization $x(t)$ this is
\begin{equation}\label{stress}
    y = \frac{1}{N-1} u^i x_i \left( x_i -\frac{1}{N} u^j x_j \right),
\end{equation}
where $u\equiv \{1\}_{i=1,...,N}$ induces the summations, and from now on we omit the time dependence for ease of notation. The motivation for this definition is that when $y$ gets large then some banks are comparatively much riskier than other banks, and this could trigger market panic and a bank run.
And while central banks and regulators can intervene on the overall market level with interest rates hikes, new legislation, or even market interventions like the TARP in the 2008 financial crisis \cite{hull2003options,calomiris2015assessment}, it is way more difficult to orchestrate the restructuring of a single bank over a short period of time.
Therefore it sounds reasonable from the regulators perspective to ask that exposures between banks, as quantified by the interaction matrix $\hat{M}$, do not destabilize the financial system by increasing the stress expectation.
Please note that an alternative definition of stress as the market level uncertainty would not affect the main result.

\paragraph*{The stress expectation.}
To compute the stress expectation $\mathbb{E}_{\hat{M}} y \equiv \mathbb{E} \left[ y \Big{|} \hat{M} \right]$ conditional on the interaction matrix $\hat{M}$ we evaluate the correlations
\begin{multline}\label{four terms}
    \frac{1}{\sigma^2}\mathbb{E}_{\hat{M}}\left[ x_i x_j \right] = \mathbb{E}_{\hat{M}}\left[ W_i W_j \right] +\beta\gamma^2 \mathbb{E}_{\hat{M}}\left[ S_i S_j \right]\\+\sqrt{\beta}\gamma \left( \mathbb{E}_{\hat{M}}\left[ W_i S_j \right]+\mathbb{E}_{\hat{M}}\left[ W_j S_i \right] \right).
\end{multline}
The Brownian motion correlations \cite{risken1996fokker,shreve2004stochastic} are simply $\mathbb{E}\left[ W_i W_j \right]=\delta_{ij} t$, and we find the other terms are
\begin{equation}\label{cross term}
    \mathbb{E}_{\hat{M}}\left[ W_i S_j \right] = \int_0^t \int_0^{t'} dt'dt'' M_{jk} {\left( e^{-\hat{A}(t'-t'')}\right)^k}_i  ,
\end{equation}
\begin{multline}\label{interaction term}
    \mathbb{E}_{\hat{M}}\left[ S_i S_j \right] 
    = \int_0^t  \int_0^{t'} \int_{t''}^t dt' dt'' dt''' M_{ik} M_{jm} \\
     \cdot \left( e^{-\hat{A}(t'-t'')}\right)^{kl}   {\left( e^{-\hat{A}(t'''-t'')}\right)^{m}}_l ,
\end{multline}
see the Supplementary Materials (SM) for full derivations.

\paragraph*{On the short-medium term.}

In a real financial network, exposures as defined by the matrix $\hat{M}$ are time-dependent, both as a result of the mark-to-market delta variations and for the booking of new trades \cite{hull2003options}.
Here we assume that exposures can be considered approximately fixed on the short-medium term, and we study the problem in an expansion. Considering only the first order for the matrix exponentials
\begin{equation}
   \left( e^{-\hat{A}\tau}\right)_{ij} = \delta_{ij} - A_{ij}\tau +\mathcal{O}(\tau^2), 
\end{equation}
and applying to Eqs. \eqref{cross term}-\eqref{interaction term} we obtain, up to $\mathcal{O}(t^3)$,
\begin{equation}
    \mathbb{E}_{\hat{M}}\left[ W_i S_j \right] = \frac{t^2}{2}\left[ M_{ji} \left(1 -\beta \frac{t}{3}\right) +\sqrt{\beta} \gamma M_{jk} {M^k}_i \frac{t}{3} \right],
\end{equation}
\begin{equation}
    \mathbb{E}_{\hat{M}}\left[ S_i S_j \right] 
    = {M_i}^k M_{jk} \frac{t^3}{3}.
\end{equation}
The stress expectation is then
\begin{multline}\label{stress expectation}
    \mathbb{E}_{\hat{M}}y = \sigma^2 t  \,+\frac{\sigma^2\sqrt{\beta}\gamma}{N-1} u^i \left( M_{ii} -\frac{1}{N} u^j M_{ij} \right) \left(1-\frac{\beta}{3} t \right) t^2  \\
    + \frac{\sigma^2 \beta\gamma^2}{3(N-1)} M_{ik} \left[ \widetilde{M}^{ik} -\frac{1}{N} u^i u_j  \widetilde{M}^{jk} \right] t^3 ,
\end{multline}
where by $\widetilde{M}_{ij}\equiv M_{ij}+M_{ji}$ we denote the symmetrized interaction matrix elements.
Let us note that even if we would start with some initial offset $x(0)\neq 0$ the expected stress change $\mathbb{E}_{\hat{M}}y(t)-y(0)$ would still be given by Eq. \eqref{stress expectation} as long as $h(0)=0$.

\paragraph*{Physical interpretation of Eq. \eqref{stress expectation}.}

The stress expectation establishes the conditions on $\hat{M}$ under which interactions are beneficial to stabilize the financial system, and in Eq. \eqref{stress expectation} we derived the leading terms in the shorter timeframe and discuss their meaning here.

The first order term $\sigma^2 t$ is the standard statistics coming from the uncorrelated Brownian motions \cite{shreve2004stochastic,risken1996fokker}.
The effect of direct interactions appears at the second order where a negative correction occurs if off-diagonal terms are overall larger than diagonal terms, meaning when $u^i M_{ii} < u^i u^j M_{ij} /N $. For the financial network this means that positive exposures reduce the expectation of stress, as indeed positive correlations imply more homogeneous returns over the banks. This effect is reduced at the third order as the noise integration time $\beta^{-1}$ is approached.

\begin{figure}
    \centering
    \includegraphics[scale=0.5]{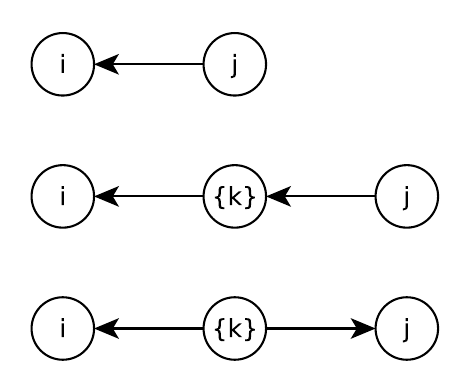}
    \caption{Diagrams contributing to the stress expectation $\mathbb{E}y(t)$ in Eq. \eqref{stress expectation} for each nodes pair $(i,j)$. The arrows denote the noise propagation and $\{ k \}$ the set of all nodes. Top) $\gamma$ order terms $M_{ij}$; Middle) $\gamma^2$ order terms $M_{ik} {M^{k}}_j$; Bottom) $\gamma^2$ order terms $M_{ik} {M_j}^k$.}
    \label{fig:one}
\end{figure}

The third order term in the second line is the effect of indirect interactions occurring in the two forms listed below, also see Fig. \ref{fig:one},
\begin{itemize}
\item the noise on node $j$ propagates to the other nodes $\{k\}$ and then on to node $i$, giving the term $u^i u_j M_{ik}M^{kj}$.
\item the noise on the other nodes $\{k\}$ affects directly both nodes $i$ and $j$ thereby creating a correlation, giving the term $u^i u_j M_{ik}M^{jk}$.
\end{itemize}
The sum of these two terms gives the symmetrized interaction matrix term $u^i u_j M_{ik}\widetilde{M}^{jk}$.

\paragraph*{Random interaction matrix.}
Assume now that regulators do not have detailed knowledge of exposures between banks, but only of the average squared exposure in the network $\gamma^2$, so that we can take $\hat{M}$ to be a random matrix with independent Gaussian entries, namely $\mathbb{E}M_{ij}=0$ and $\mathbb{E}\left[ M_{ij} M_{kl} \right] = \delta_{ik}\delta_{jl}$.
The random matrix expectation of the expected stress is, up to $\mathcal{O}(t^3)$,
\begin{equation}\label{EEy}
    \mathbb{E}y = \sigma^2 t \left[ 1 + \frac{\beta \gamma^2}{3} (N+1) \, t^2 \right],
\end{equation}
and we note that stochastic dynamics and random matrix expectations commute.
This Eq. \eqref{EEy} is consistent with the fact that adding interactions in an unknown direction is effectively an additional source of noise, which on average can only increase the sample variance $y$.

\begin{figure}\label{two}
    \centering
    \includegraphics[scale=0.55]{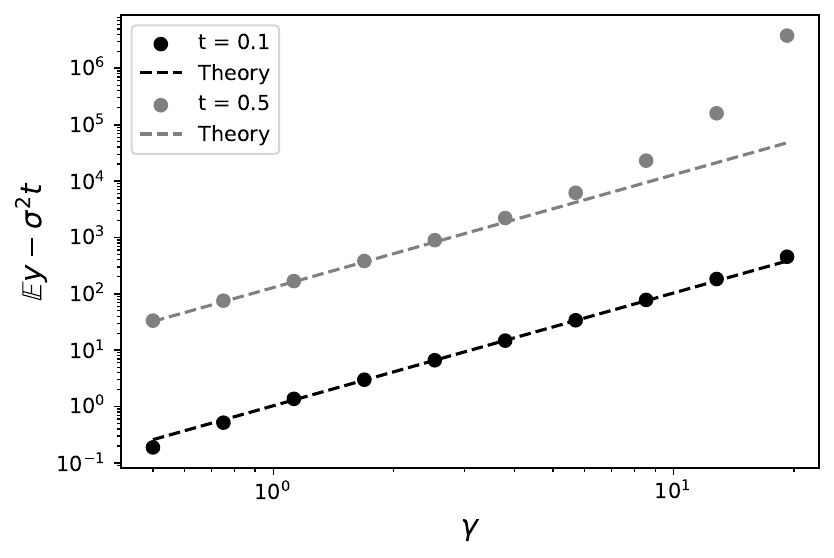}
    \caption{Quadratic scaling of the interaction correction with the interaction strength $\gamma$ according to the theoretical estimate of Eq. \eqref{EEy}. The parameters used here are $N=30$, $\sigma = 100$, $\beta=0.01$.}
    \label{fig:two}
\end{figure}

The interaction correction of Eq. \eqref{EEy} results to be a lower bound for larger time intervals, see the numerical results in Fig. \ref{fig:two}. This is understood as the random matrix $\hat{A}$ has a positive probability of having the highest eigenvalue larger than zero which makes the Ornstein-Uhlenbeck process non stationary and expectations diverging exponentially. 

The variance of the interaction correction on the stress expectation due to the uncertainty about the interaction matrix is, up to $\mathcal{O}(t^4)$, see SM,
\begin{equation}\label{correction variance}
    \mathbb{E} \left[ \left( \mathbb{E}_{\hat{M}} y    - \mathbb{E} y  \right)^2  \right]   = \sigma^4 \frac{\beta \gamma^2}{N-1} \,t^4, 
\end{equation}
so that the standard deviation is of lower order than the interaction correction of Eq. \eqref{EEy}.
This means that, while interactions on average destabilize the dynamics, a bigger impact comes actually from the uncertainty about these interactions, and especially for small networks, which motivates the importance of disclosure regarding exposures between big banks.

\paragraph*{Network size.}

The network size effect is described by the linear factor $N+1$ in the interaction correction of Eq. \eqref{EEy} and in the $1/(N-1)$ factor in the variance of Eq. \eqref{correction variance}, which implies that for large networks the destabilization becomes certain. However, in obtaining these results we did not consider any constraints on the total exposures, while actually capital regulations are in place in financial markets requiring banks to limit their risk taking. Assuming a simple statistical constraint on exposures of the form $\mathbb{E} u^j (\gamma M_{ij})^2 = k^2$ with $k$ independent of $N$ gives $\gamma = k / \sqrt{N}$ so that $\mathbb{E}y$ is not increasing with the network size. An even more strict constraint of the form $\mathbb{E} u^j \gamma \big| M_{ij} \big| = k$ would give $\gamma = \sqrt{\frac{\pi}{2}} \frac{k}{N}$ and $\mathbb{E}y-\sigma^2 t$ converging to zero for large $N$ in agreement with the law of large numbers.
Let us also note that imposing a no self-interaction condition, meaning $M_{ii}=0$ for all $i$s, it does not change Eq. \eqref{EEy}.


\paragraph*{Stochastic volatility.}
To motivate the consideration of the interaction correction we compare it with the impact of stochastic volatility which is an industry standard for the pricing models of interest rates derivatives \cite{shreve2004stochastic,brigo2006interest}. Modeling volatility as a geometric Brownian motion with VolVol parameter $\nu$, we find that corrections to the correlations $\mathbb{E}\left[ x_i x_j \right] $ appear already at $t^2$ order but at this order they are independent of the indices $i$ and $j$, so the stochastic volatility correction to Eq. \eqref{stress expectation} appears only at $\mathcal{O}(t^3)$, see SM,
\begin{equation}\label{volatility correction}
    \mathbb{E}_{\hat{M}}y - \mathbb{E}_{\hat{M}}y^{\nu=0} = \sigma^2 \sqrt{\beta}\gamma \nu^2 u^i \left( \widetilde{M}_{ii}  -\frac{1}{N} u^j \widetilde{M}_{ij}  \right)\frac{t^3}{6},
\end{equation}
and it is dominated by the direct interaction correction on the short time limit.
This stochastic volatility correction of Eq. \eqref{volatility correction} vanishes with the random matrix expectation as $\mathbb{E}\widetilde{M}_{ij} = 0$, so that Eq. \eqref{EEy} stays unaffected.

\begin{figure}\label{three}
    \centering
    \includegraphics[scale=0.55]{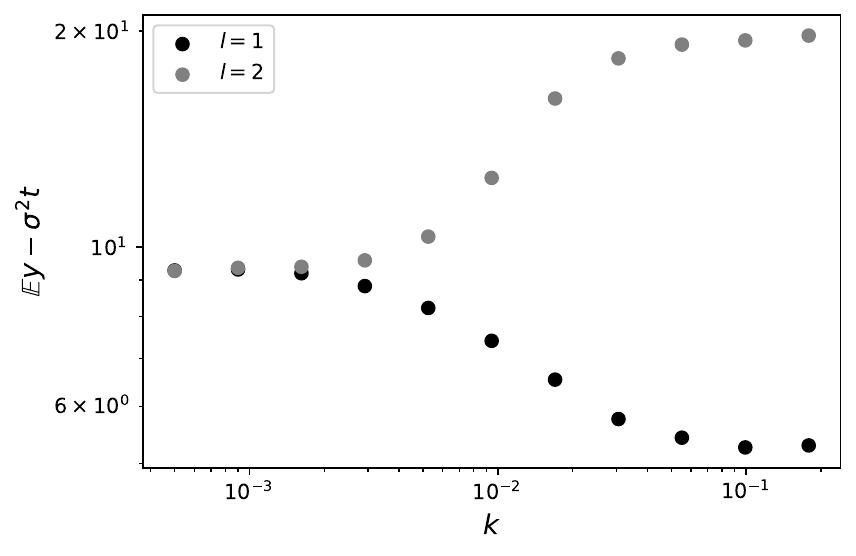}
    \caption{Effect of nonlinearities. $k$ quantifies the dependence of the bond's delta on the spread, and $l$ the maximum increase factor in the exposure. The parameters used here are $N=30$, $\sigma = 100$, $\gamma = 3$, $\beta=0.01$, and $T=0.1$.}
    \label{fig:three}
\end{figure}

\paragraph*{Effect of nonlinearities.}
The assumption of a constant interaction matrix is valid only for short timescales as it neglects the time variation of portfolios' delta risks, which is due to trades' time to maturity getting shorter, interest rates and hazard rate curves changes, new trades being booked, and more \cite{shreve2004stochastic,brigo2006interest}. These effects introduce a number of nonlinearities which impact the stress expectation.
As an example, consider zero-coupon risky bonds \cite{oKane2011modelling} whose present value can be modelled in the simplest case as $B_j(T)=\exp \left[ - (r+x_j) T \right]$, where $T$ is the time to maturity, $r$ is some constant interest rate, and $x_j$ is the market-implied spread of reference entity $j$.
We see that the exposure to variations of the spread $x_j$ is not a constant, $\partial_{x_j}B_j(T) = -T B_j(T)$, as it depends on both $T$ and $x_j$.
Note that in our framework the impact on $i$'s balance sheet due to exposures to $j$'s bonds will affect the hazard rate $x_i$ over a timescale $\beta^{-1}$ through the recent history integral.
We study the effect of this simple bond nonlinearity in an approximation by introducing the factor $M_{ij}\rightarrow M_{ij}\min[\exp{(-k x_j(t))},l]$, with $k\geq 0$ and $l\geq 1$ parameters, where $l$ is to upper bound the bond price to roughly its notional value.
Numerically it is found that in this nonlinear example the interaction correction can be smaller or greater than in the linear case, see Fig. \ref{fig:three}, as indeed the convexity effect governed by $k$ will increase exposures and on average destabilize the system, while the upper bound on exposures given by the maximum factor $l$ will accordingly contain it.

\paragraph*{Eigenvalues.}
If the interaction matrix $\hat{M}$ is such that the largest eigenvalue of the matrix $\hat{A}$ is larger than zero, then the recent history dynamics $h$ is not stationary and the network state $x$ will diverge exponentially in the long term. 
On shorter timescales, however, we see from the interaction correction of Eq. \eqref{stress expectation} that the dynamics depends on other properties of the interaction matrix, and we may ask if these are related to eigenvalues. Take for simplicity $-\gamma=\beta^{-\frac{1}{2}}\gg 1$ so that $\hat{A}\approx \hat{M}$ and we can directly study the eigenvalues of $\hat{M}$. With numerical simulations we find that the average eigenvalue of $\hat{M}$, which indeed is equivalent to the trace $u^i M_{ii}$, is correlated to $u^i \left( M_{ii} -\frac{1}{N} u^j M_{ij} \right) $ which is the $\gamma$ order term in Eq. \eqref{stress expectation}. Then we find that the sample variance of the eigenvalues is correlated to $M_{ik} \left[ \widetilde{M}^{ik} -\frac{1}{N} u^i u_j  \widetilde{M}^{jk} \right] $ which is the $\gamma^2$ order term in Eq. \eqref{stress expectation}, see SM.
These results suggest that not only the eigenvector of the largest eigenvalue but all eigenvectors play a role on the short-medium timescale.

\paragraph*{Discussion.}
In this Letter, based on the definition of a financial network stress observable in the perspective of regulators, an analytical framework to estimate the short-medium term impact of interbank exposures has been introduced, to complement the existing methods concerning the long term stability of complex systems \cite{may1972will,haldane2011systemic,barzel2013universality,bardoscia2017pathways,meena2023emergent,bontorin2023multi}.
This short-medium term characterization is particularly relevant when the interaction matrix describing exposures is time-dependent.
The focus here is on the interbank liabilities only since the effect of overlapping portfolios is already well-studied \cite{shreve2004stochastic}, and would enter the model as a correlation between noise components.
Also note that defaults are not explicitly considered based on the assumption that most of the bondholders losses will happen as fair price variations in the weeks and months before the default event.

In the proposed framework, the expected stress level was derived as a function of the interaction matrix, and to leading order it is found that positive exposures have a stabilizing effect as indeed they are levelling out returns relative to the market average.
Then it is shown that, if exposures are not exactly known but only their expected strength is, then this effectively works as an additional noise source to increase the stress expectation following a quadratic law, and even bigger is shown to be the impact of the uncertainty about single realizations.
The effect of nonlinearities and of stochastic volatility are also studied for simple examples. These results suggest the importance of disclosing interbank exposures in the perspective of financial regulators.





\bibliography{apssamp}

\end{document}


\preprint{APS/123-QED}

\title{Supplementary Materials for the manuscript "Interaction uncertainty in financial networks"}

\author{Andrea Auconi}

\affiliation{%
Ca’ Foscari University of Venice, DSMN , via Torino 155, 30172 Mestre (Venice), Italy
}%


\date{\today}



\maketitle


\section{Derivation of Eqs. (7)-(8)}
Consider Eq. (3) in the main text, $x_i(t)= \sigma W_i(t)+\sigma\sqrt{\beta}\gamma S_i(t)$.
We start by reviewing the standard Brownian motion correlations \cite{karatzas1991brownian,shreve2004stochastic} in our formalism,
\begin{multline}
    \mathbb{E}\left[ W_i(t) W_j(t) \right] 
    \\= \mathbb{E} \left[\int_0^t \int_0^t  dW_i(t') dW_j(t'') \right] 
    = \int_0^t \int_0^t \mathbb{E}\left[ dW_i(t') dW_j(t'') \right] \,
    =  \int_0^t \int_0^t \delta_{ij} dt' \delta_{t't''}
    =\delta_{ij} \int_0^t  dt'\, \mathbb{I}_{0<t'<t} \, = \delta_{ij} t,
\end{multline}
where we used the delta definition $\int_0^t \delta_{t't''} f(t'') = f(t')\, \mathbb{I}_{0<t'<t}$.
Similarly we derive Eq. (6) of the main text as
\begin{multline}
    \mathbb{E}_{\hat{M}}\left[ W_i(t) S_j(t) \right] \\
    = \int_0^t dt' \int_0^{t'} M_{jk} \left( e^{-\hat{A}(t'-t'')}\right)^{kl} \mathbb{E}\left[ dW_l (t'') \int_0^t dW_i(t''')  \right] \\
    =\int_0^t dt' \int_0^{t'} M_{jk} \left( e^{-\hat{A}(t'-t'')}\right)^{kl} \delta_{li} dt'' \int_0^t \delta_{t''t'''}
    \, =\int_0^t dt' \int_0^{t'} M_{jk} {\left( e^{-\hat{A}(t'-t'')}\right)^k}_i  dt''.
\end{multline}
For Eq. (7) of the main text we have
\begin{multline}
    \mathbb{E}_{\hat{M}}\left[ S_i(t) S_j(t) \right]\\
    = \int_0^t dt' \int_0^{t'} M_{ik} \left( e^{-\hat{A}(t'-t'')}\right)^{kl} 
    \int_0^t dt''' M_{jm}  \mathbb{E}_{\hat{M}}\left[ dW_l (t'') \int_0^{t'''} \left( e^{-\hat{A}(t'''-t'''')}\right)^{mq} dW_q(t'''')  \right] \\
    = \int_0^t dt' \int_0^{t'} M_{ik} \left( e^{-\hat{A}(t'-t'')}\right)^{kl} 
    \int_0^t dt''' M_{jm} dt'' \int_0^{t'''} \left( e^{-\hat{A}(t'''-t'''')}\right)^{mq} \delta_{lq} \delta_{t''t''''}  \\
    = \int_0^t dt' \int_0^{t'} M_{ik} \left( e^{-\hat{A}(t'-t'')}\right)^{kl} 
    \int_0^t dt''' M_{jm}  dt'' \left( e^{-\hat{A}(t'''-t'')}\right)^{mq} \delta_{lq} \mathbb{I}_{t''<t'''} \\
    = \int_0^t dt' \int_0^{t'} dt'' M_{ik} \left( e^{-\hat{A}(t'-t'')}\right)^{kl} 
    \int_{t''}^t dt''' M_{jm}  {\left( e^{-\hat{A}(t'''-t'')}\right)^{m}}_l .
\end{multline}

\section{Derivation of Eqs. (13) - (14)}
Recall the random matrix expectation definitions $\mathbb{E}M_{ij}=0$ and $\mathbb{E}\left[ M_{ij} M_{kl} \right] = \delta_{ik}\delta_{jl}$. Considering the stress expectation equation (Eq. (12) in the main text), for the random matrix expectation $\mathbb{E}y(t)$ it suffices to evaluate
\begin{multline}
    \mathbb{E}\left[ M_{ik} \widetilde{M}^{ik} \right] = \mathbb{E}\left[ M_{ik} M^{ik} \right]  + \mathbb{E}\left[ M_{ik} M^{ki} \right] \\
    = u^i u^k \mathbb{E}\left[ (M_{ik})^2 \right]  + u^i u^k\mathbb{E}\left[ M_{ik} M_{ki} \right] 
    = u^i u^k u_i u_k + u^i u^k \delta_{ik}
    = N^2 + N;
\end{multline}
\begin{multline}
   u^i u_j \mathbb{E}\left[ M_{ik} \widetilde{M}^{jk} \right] 
   = u^i u^j u^k \mathbb{E}\left[ M_{ik} M_{jk} \right] + u^i u^j u^k \mathbb{E}\left[ M_{ik} M_{kj} \right] \\
   = u^i u^j u^k \delta_{ij} \delta_{kk} + u^i u^j u^k \delta_{ik} \delta_{kj}
   = u^i u^j u^k \delta_{ij} \delta_{kk} + u^i u^j u^k \delta_{ik} \delta_{kj} 
   = N^2 + N.
\end{multline}

Similarly for deriving Eq. (14) of the main text we evaluate
\begin{equation}
    \mathbb{E} \left[ \left( \mathbb{E}_{\hat{M}} y    - \mathbb{E} y  \right)^2  \right] 
    = \sigma^4 \frac{\beta \gamma^2}{(N-1)^2} t^4 \,\mathbb{E} \left[  \left( u^i \left(M_{ii} -\frac{1}{N}  u^j M_{ij} \right) \right)^2 \right]+ \mathcal{O}(t^5) 
    = \sigma^4 \frac{\beta \gamma^2}{N-1} t^4 + \mathcal{O}(t^5), 
\end{equation}
where we computed
\begin{multline}
    \mathbb{E} \left[  \left( u^i \left(M_{ii} -\frac{1}{N}  u^j M_{ij} \right) \right)^2 \right] 
    = u^i u^k \mathbb{E}\left[ M_{ii} M_{kk}  \right] -\frac{1}{N} u^i u^k u^l \mathbb{E}\left[ M_{ii} M_{kl}  \right]
    -\frac{1}{N} u^i u^j u^k \mathbb{E}\left[ M_{ij} M_{kk}  \right] + \frac{1}{N^2} u^i u^j u^k u^l \mathbb{E}\left[ M_{ij} M_{kl}  \right] \\
    = u^i u^k  \delta_{ik} -\frac{1}{N} u^i u^k u^l \delta_{ik} \delta_{il} 
    -\frac{1}{N} u^i u^j u^k \delta_{ik} \delta_{jk} + \frac{1}{N^2} u^i u^j u^k u^l \delta_{ik} \delta_{jl} \,
    = N - 1.
\end{multline}

\section{Derivation of Eq. (15)}
The stochastic volatility process is here modeled as a geometric Brownian motion, $d\sigma(t) = \nu \sigma(t) dB$, where $dB$ are standard Brownian motion increments independent of $dW$. The solution is well-known,
\begin{equation}
    \sigma(t) = \sigma \exp \left[ \nu B(t) -\frac{\nu^2}{2} t \right],
\end{equation}
and it can be found using It\^o's Lemma \cite{karatzas1991brownian,shreve2004stochastic}, and we denoted for simplicity $\sigma\equiv \sigma(0)$.

Let us define the stochastic volatility term
\begin{equation}
    \xi(t) \equiv \exp \left[ \nu B(t) -\frac{\nu^2}{2} t \right]  -1 ,
\end{equation}
and rewrite our financial network dynamics model (Eq. (1) in the main text) including it, 
\begin{equation}
   dx_i = \sigma \left( 1+\xi \right) dW_i + \gamma M_{ij} h^j dt, 
\end{equation}
so that the formal solution is
\begin{equation}
    x_i(t) = x_i^{\nu = 0}(t) + \sigma \int_0^t \xi(t') dW_i(t') +\sigma \sqrt{\beta}\gamma \int_0^t dt' \int_0^{t'} M_{ij} \left( e^{-\hat{A}(t'-t'')}\right)^{jk} \xi(t'') dW_k (t''),
\end{equation}
where $x_i^{\nu = 0}(t)$ is the solution with constant volatility $\nu=0$ and the same realization of $\{ W(t') \}_{t'\in [0,t]}$.
Given that $\mathbb{E}\xi(t)=0$ as the process is a martingale \cite{karatzas1991brownian,shreve2004stochastic}, the expectation $\mathbb{E}_{\hat{M}}\left[ x_i(t) x_j(t) \right]-\mathbb{E}\left[ x_i(t) x_j(t) \right]^{\nu=0}$ do not involve cross terms with $x_i^{\nu = 0}(t)$,
\begin{multline}\label{vol_calc1}
    \mathbb{E}_{\hat{M}}\left[ x_i(t) x_j(t) \right] = \mathbb{E}_{\hat{M}}\left[ x_i(t) x_j(t) \right]^{\nu=0}
    + \sigma^2 \int_0^t \int_0^t \mathbb{E}\left[\xi(t')\xi(t'') dW_i(t') dW_j(t'')\right]\\
    + \sigma^2 \sqrt{\beta}\gamma \int_0^t dt' \int_0^{t'} M_{ik} \left( e^{-\hat{A}(t'-t'')}\right)^{kl} \mathbb{E}\left[ \int_0^t \xi(t'')\xi(t''') dW_l (t'')dW_j (t''') \right] \\
    + \sigma^2 \sqrt{\beta}\gamma \int_0^t dt' \int_0^{t'} M_{jk} \left( e^{-\hat{A}(t'-t'')}\right)^{kl} \mathbb{E}\left[ \int_0^t \xi(t'')\xi(t''') dW_l (t'')dW_i (t''') \right]
    \,+\mathcal{O}(t^4)\\
    = \sigma^2\frac{\nu^2}{2} t^2 +\sigma^2\frac{\nu^4}{6} t^3 + \sigma^2 \sqrt{\beta}\gamma \nu^2 \widetilde{M}_{ij} \frac{t^3}{6} +\mathcal{O}(t^4),
\end{multline}
where we computed
\begin{multline}
    \int_0^t \int_0^t \mathbb{E}\left[\xi(t')\xi(t'') dW_i(t') dW_j(t'')\right] \,
    = \int_0^t \int_0^t \mathbb{E}\left[\xi(t')\xi(t'') \right] \mathbb{E}\left[dW_i(t') dW_j(t'')\right] \,
    = \int_0^t \int_0^t \mathbb{E}\left[\xi(t')\xi(t'') \right] \delta_{t't''} dt' \\
    = \int_0^t dt' \,\mathbb{E}\left[\xi^2(t')\right] \,
    = \int_0^t dt'  \left( e^{\nu^2 t'} \mathbb{E}\left[ e^{(2\nu) B(t') -\frac{(2\nu)^2}{2} t' } \right]    -1 \right)\,
    = \int_0^t dt' \left( e^{\nu^2 t'}  -1 \right)\,
    = \frac{\nu^2}{2} t^2 +\frac{\nu^4}{6} t^3 \,+\mathcal{O}(t^4).
\end{multline}
and also
\begin{multline}
    \int_0^t dt' \int_0^{t'} M_{ik} \left( e^{-\hat{A}(t'-t'')}\right)^{kl} \mathbb{E}\left[ \int_0^t \xi(t'')\xi(t''') dW_l (t'')dW_j (t''') \right]\\
    =\int_0^t dt' \int_0^{t'} M_{ik} \left( e^{-\hat{A}(t'-t'')}\right)^{kl} \delta_{lj} dt'' \int_0^t \delta_{t''t'''}  \mathbb{E}\left[\xi(t'')\xi(t''')\right] \\
    =\int_0^t dt' \int_0^{t'} dt'' M_{ik} \left( e^{-\hat{A}(t'-t'')}\right)^{kl} \delta_{lj} \mathbb{E}\left[\xi^2(t'') \right] \,
    =\int_0^t dt' \int_0^{t'} dt'' M_{ik} {\left( e^{-\hat{A}(t'-t'')}\right)^k}_j \mathbb{E}\left[\xi^2(t'') \right] \\
    =\int_0^t dt' \int_0^{t'} dt'' M_{ik} {\left( e^{-\hat{A}(t'-t'')}\right)^k}_j \left( e^{\nu^2 t''} -1 \right) \,
    =\nu^2 \int_0^t dt' \int_0^{t'} dt'' M_{ik} {\delta^k}_j t''  + \mathcal{O}(t^4)\\
   = \nu^2 M_{ij} \frac{t^3}{6}  + \mathcal{O}(t^4) .
\end{multline}

We see that the first and second terms in Eq. \eqref{vol_calc1} are independent of the indices $i$ and $j$ so they have no contribution to $y(t)$ as it is immediately seen from the definition of Eq. (5) of the main text. Then we obtain the stochastic volatility correction to the stress expectation as
\begin{equation}
    \mathbb{E}y(t) = \mathbb{E}y(t)^{\nu=0}\\ + \sigma^2 \sqrt{\beta}\gamma \nu^2 u^i \left( \widetilde{M}_{ii}  -\frac{1}{N} u^j \widetilde{M}_{ij}  \right)\frac{t^3}{6} \,+ \mathcal{O}(t^4).
\end{equation}

\section{Eigenvalues}
\begin{figure}
    \centering
    \includegraphics[scale=0.45]{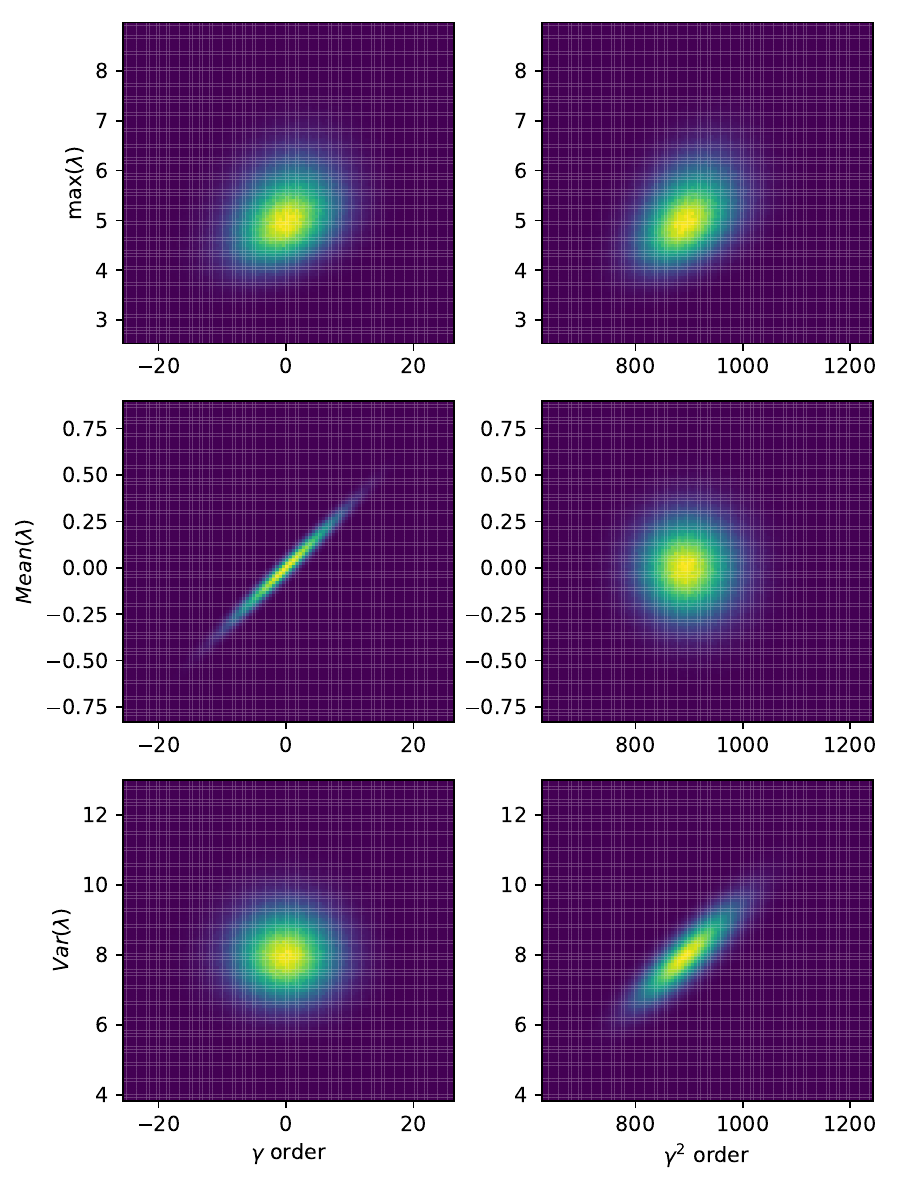}
    \caption{Correlation of the expansion terms of Eq. (12) of the main text with properties of the interaction matrix eigenvalues $\{\lambda\}$, with network size $N=30$.}
    \label{fig:A1}
\end{figure}

The discussion of eigenvalues in the main text is here complemented with a plot. Given the set of eigenvalues $\{\lambda\}$ of the interaction matrix $\hat{M}$, we see that the first and second order in the interaction strength $\gamma$ of the stress expectation expansion of Eq. (12) of the main text are correlated respectively with the mean $\bar{\lambda}=\frac{1}{N}\sum_{i=1}^N \lambda_i$ and variance $Var(\lambda)=\frac{1}{N}\sum_{i=1}^N (\lambda_i-\bar{\lambda})^2$, see Fig. A\ref{fig:A1}.

\section{Market level uncertainty}\label{Appendix F}
Defining the market level as the sample mean over the banks, then the market level uncertainty $z(t)$ is
\begin{multline}
    z(t) \equiv \mathbb{E} \left[ \left( \frac{1}{N} u^i x_i(t) \right)^2 \right]\, =    \frac{1}{N^2} u^i u^j \mathbb{E} \left[ x_i(t) x_j(t)  \right]\\
    = \sigma^2 t  +\frac{\sigma^2\sqrt{\beta}\gamma}{N^2} u^i u^j M_{ij}  \left(1-\frac{\beta}{3} t \right) t^2  \,
    + \frac{\sigma^2 \beta\gamma^2}{3 N^2}  u^i u_j M_{ik} \widetilde{M}^{jk}  t^3 ,
\end{multline}
since $\mathbb{E}\left[ \frac{1}{N} u^i x_i(t) \right] =0$. We see that to second order $\mathcal{O}(t^2)$ the effect of off-diagonal terms is opposite to $z(t)$ compared to $y(t)$. However, the interaction correction in the random matrix expectation is destabilizing also for the market level uncertainty,
\begin{equation}
    \mathbb{E} z(t) = \sigma^2 t  + \frac{\sigma^2 \beta\gamma^2}{3}  \left(1+\frac{1}{N} \right)  t^3.
\end{equation}

\bibliography{apssamp}